\documentclass[11pt]{article}
\usepackage[margin=1in]{geometry}
\usepackage{amsmath, amssymb}
\usepackage{graphicx}
\usepackage{hyperref}
\usepackage{pslatex}
\usepackage{apacite}
\usepackage{float} 
\usepackage{algorithm}
\usepackage{algorithmic}
\usepackage{booktabs}
 
\title{When not to help: planning for lasting human-AI collaboration}

\author{
  {\large \bf Mark Steyvers} \\
  Department of Cognitive Sciences \\
  University of California, Irvine \\
  \texttt{mark.steyvers@uci.edu}
  \and
  {\large \bf Lukas Mayer} \\
  Department of Cognitive Sciences \\
  University of California, Irvine \\
  \texttt{lwmayer@uci.edu}
}

\date{}

\begin{document}
\maketitle

\begin{abstract}
AI systems and technologies that can interact with humans in real time face a communication dilemma: when to offer assistance and how frequently. Overly frequent or contextually redundant assistance can cause users to disengage, undermining the long-term benefits of AI assistance. We introduce a cognitive modeling framework based on Partially Observable Markov Decision Processes (POMDPs) that addresses this timing challenge by inferring a user’s latent cognitive state related to AI engagement over time. Additionally, our framework incorporates reasoning about the long-term effects of AI assistance, explicitly aiming to avoid actions that could lead the human user to disengage or deactivate the AI. A key component of our approach is counterfactual reasoning: at each time step, the AI considers how well the user would perform independently and weighs the potential boost in performance against the risk of diminishing engagement with the AI. Through simulations, we show that this adaptive strategy significantly outperforms baseline policies in which assistance is always provided or never provided. Our results highlight the importance of balancing short-term decision accuracy with sustained user engagement, showing how communication strategies can be optimized to avoid ``alert fatigue'' while preserving the user’s receptiveness to AI guidance.

\end{abstract}

\section{Introduction}
Modern AI systems and technologies are increasingly integrated into our daily lives, offering guidance, alerts, and assistance in a wide range of contexts. However, their success hinges on timely communication—namely, knowing when to convey important information and when to remain silent. A common and often frustrating issue arises when these systems provide excessive or redundant advice, failing to account for the user’s familiarity with a situation. For example, consider a driver pulling into their home garage—a task they have completed successfully countless times—only to be inundated with warnings about potential collisions from the car’s assistance system. Such alerts, while well-intentioned, may feel unnecessary and intrusive, leading the user to disregard them entirely. Over time, this can result in “alert fatigue,” where the user becomes desensitized to critical warnings due to frequent exposure to non-essential ones \cite{ancker2017effects}.

Similarly, AI systems designed to assist in decision-making often fail to adapt to a user’s level of expertise or familiarity with a task. For instance, an AI advisor providing investment recommendations might over-explain basic financial principles to a seasoned investor, reducing the perceived value of its assistance. Over time, the user may disengage, ignoring even high-quality advice in unfamiliar or complex scenarios where the AI’s guidance could have been beneficial. These examples highlight the dual challenge faced by AI systems: providing timely and useful assistance while avoiding over-advising that can diminish user engagement. As emphasized in general guidelines for human-AI interaction \cite{amershi2019guidelines}, achieving this balance requires careful consideration of the timing and frequency of AI advice, ensuring that assistance is delivered only when it is contextually relevant and beneficial.

Addressing this challenge requires an understanding of how users interact with AI systems and how their engagement evolves over time. Foundational work in AI assistance, such as the seminal frameworks proposed by \cite{horvitz1999principles,horvitz2003models}, introduced utility-based approaches to balance the benefits of AI intervention against its potential costs. These frameworks involve formulating the decision to provide assistance as an optimization problem, where the AI maximizes an expected utility function that accounts for factors such as task outcomes and the cost of AI intervention. While these approaches have been instrumental in advancing decision-theoretic AI assistance, they require an explicit formulation of the penalty of over-advising.

In contrast, the computational framework proposed in this paper shifts the focus to a reasoning problem about the human’s cognitive state—specifically, their engagement and adherence to AI advice. This framework accounts for how adherence evolves over time as a function of the perceived usefulness of past advice. By incorporating a cognitive model, the AI can predict the likelihood of the human making the correct decision both with and without AI assistance. In this framework, the cost of over-advising arises naturally as the consequence of disengagement, leading to poor decisions and missed opportunities where AI assistance could be beneficial.

This approach builds on insights from recent studies in human-AI interaction \cite{chen2024learning,noti2023learning,buccinca2024towards}, which emphasize the importance of understanding and adapting to user behavior but extends them by explicitly incorporating engagement dynamics as a key variable in the decision-making process.

In this paper, we propose a cognitive modeling framework based on Partially Observable Markov Decision Processes (POMDPs) that addresses these challenges by adapting assistance to a user's inferred cognitive state while explicitly reasoning about the long-term impact of its interventions to avoid user disengagement or system deactivation. We detail the theoretical underpinnings, provide simulation results, and discuss the potential for applying these methods. However, we note that we do not yet incorporate results from behavioral studies to test the assumptions of the cognitive modeling framework, leaving that extension for future research. 

\section{Cognitive Modeling Framework}
We present a computational framework that addresses the trade-offs in providing AI assistance by dynamically adapting to the user's engagement state. Unlike traditional utility-based approaches, which optimize static reward functions, our framework integrates a cognitive model of engagement dynamics, enabling the AI to balance immediate decision accuracy with the long-term goal of maintaining user engagement.

The core idea is to model the interaction between the AI and the human as a sequential decision-making problem. The AI must decide, at each time step, whether to provide assistance based on the human’s task context and the state of engagement, which determines the probability that the user adheres to the AI assistance. Our approach builds on the principles of Partially Observable Markov Decision Processes (POMDPs), which enable the AI to reason under uncertainty. The state space includes both observable variables, such as task context and human decisions, and latent variables, such as the human’s engagement state at each point in time. Through this framework, the AI learns to optimize its advising strategy, ensuring that assistance is provided when it is most needed while avoiding unnecessary intervention in familiar or simple contexts.

The following subsections detail the key components of this framework, including the problem setup, the modeling of human decision-making with and without AI assistance, and the dynamics of engagement.

\begin{figure}[H]
    \centering
    \includegraphics[width=0.7\linewidth,
                     trim={0cm 0.5cm 0cm 0.0cm},
                     clip]{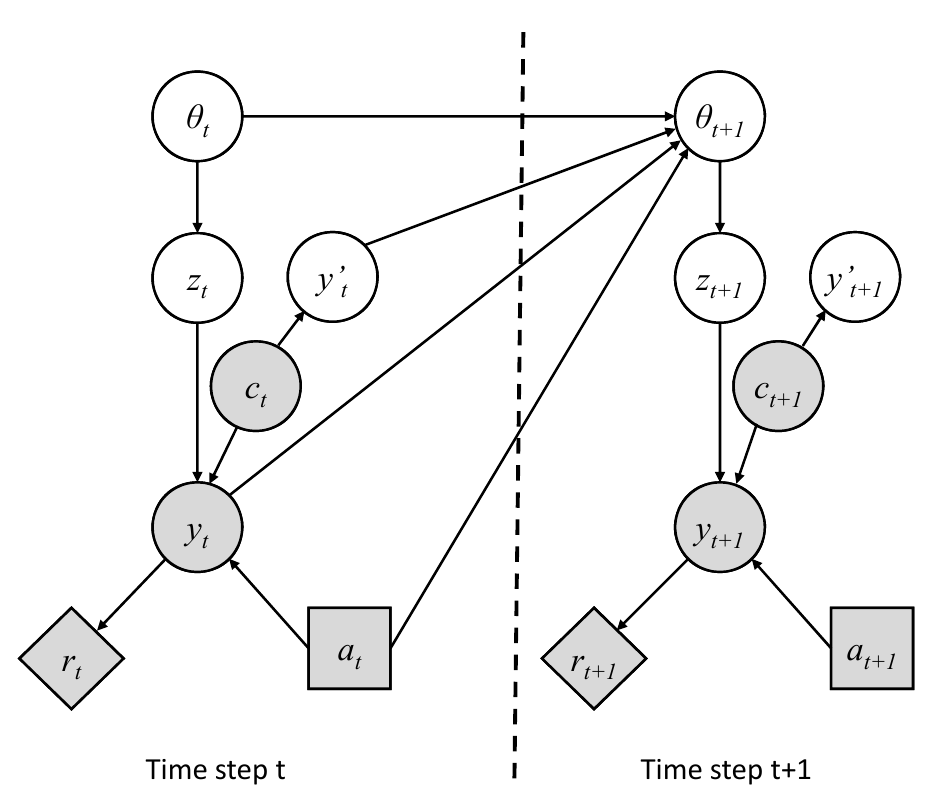}
    \caption{Dynamic graphical model that shows the dependencies between states over time. The shaded symbols represent information observable to the AI assistant. The unshaded symbols represent latent cognitive states. The variable $y_t$ is human decision with the AI assistance. The variable $y'_t$ represents a counterfactual decision had the AI assistance not occurred.}
    \label{fig:diagram}
\end{figure}

\subsection{Problem Setup}
At each time step \( t \), the human makes a binary decision \( y_t \in \{\text{correct}, \text{incorrect}\} \). The correctness of the decision is influenced by the task context \( c_t \), which varies over time and determines the difficulty of making the correct choice. We simplify the contexts into two levels of familiarity: \( \{\text{low}, \text{high}\} \). The AI can either provide advice (\( a_t = \text{on} \)) or withhold it (\( a_t = \text{off} \)).

The human's decision-making behavior is modeled under two conditions: with and without AI assistance. When advice is withheld (\( a_t = \text{off} \)), the probability of making a correct decision depends solely on the task context. In contrast, when advice is provided (\( a_t = \text{on} \)), the outcome depends on whether the human adheres to the advice and the given task context.

\paragraph{Human Decisions without AI assistance}
The probability of the human making a correct decision without AI advice depends solely on the task context:
\begin{equation}
p( y_t = correct \mid a_t = off ) = 
\begin{cases} 
\alpha_H & \text{if } c_t = high, \\
\alpha_L & \text{if } c_t = low.
\end{cases}
\label{eq:withoutai}
\end{equation}
where \( \alpha_L \) and \( \alpha_H \) are context-specific model parameters.

\paragraph{Human Decisions with AI assistance}
AI advice is assumed to be error-free, ensuring correct decisions if the human adheres to it. Let \( z_t \in \{ 0, 1 \} \) represent the human's adherence to AI advice at time step $t$. If \( z_t = 1 \), the user is paying attention to and adhering to the AI ensuring that the decision is always correct. If \( z_t = 0 \), the AI is ignored by the user and the decision is made independently. Thus, the probability of a correct decision with AI advice is:

\begin{equation}
\begin{aligned}
& p( y_t = correct \mid a_t = on ) = \\
& \quad
\begin{cases} 
    1 & \text{if } z_t = 1, \\
    p( y_t = correct \mid a_t = off ) & \text{if } z_t = 0.
\end{cases}
\end{aligned}
\label{eq:withai}
\end{equation}

\subsection{Engagement Dynamics}
The engagement state \( \theta_t \) captures the degree to which the user is engaged with the AI, and governs the likelihood that the human adheres to the AI's advice through a Bernoulli distribution: 

\begin{equation}
z_t \sim \mathrm{Bernoulli}( \theta_t )
\end{equation}

The engagement state \( \theta_t \) evolves dynamically based on the interaction outcomes between the human and the AI. Updates to \( \theta_t \) reflect discrepancies between the actual human decision when advice is provided and the counterfactual decision the human would have made without AI assistance. 

The attentional state updates are formally defined as:
\begin{equation}
\begin{aligned}
& \theta_{t+1} = \\
& 
\begin{cases} 
\theta_t ( 1 - \eta ) & \text{if } a_t=on, y_t=correct, y'_{t}=correct \\
\theta_t + (1-\theta_t) \eta & \text{if } a_t=on, z_t=1, y_t=correct, y'_{t}=incorrect \\
\theta_t + (1-\theta_t) \eta & \text{if } a_t=on, z_t=0,y_t=incorrect \\
\theta_t  & \text{if } a_t=off, y_t=correct \\
\theta_t + (1-\theta_t) \eta & \text{if } a_t=off, y_t=incorrect \\
\end{cases}
\end{aligned}
\label{eq:thetaupdate}
\end{equation}

where \( \eta \) is a parameter controlling the rate of engagement adjustment, and \( y'_t \) represents the counterfactual decision when AI assistance would not have been presented. In the model, we assume that the human is able to distinguish between the actual outcome and this counterfactual outcome.  

This formulation naturally incorporates the costs of over-advising through changes in attention. For example:
\begin{itemize}
    \item If the AI advice is redundant (\( y_t = y'_t = \text{correct} \)), attention decreases, reflecting diminished perceived usefulness.
    \item If AI advice corrects a potential error the human could have made (\( y_t = \text{correct}, y'_t = \text{incorrect} \)), attention increases, reinforcing the value of advice.
    \item If the human ignores AI advice and makes a mistake (\( y_t = \text{incorrect} \)), attention increases as the human learns to rely more on AI.
\end{itemize}

This dynamic model enables the AI to adapt its advising strategy, balancing immediate decision support with long-term engagement.

\subsection{Context Dynamics}
To complete the framework, we need to specify how the task context changes over time. These contextual changes will be modeled as a simple change process where at every time step, there is a probability \( \phi \) that the task context switches from low to high or high to low familiarity. 

\subsection{Partially Observable Markov Decision Processes (POMDPs)}
The cognitive modeling approach we have introduced can be framed as a Partially Observable Markov Decision Process (POMDP) where the AI timing strategy is a sequential decision-making process under uncertainty. At each time step, the AI observes the task context as well as the correctness of each human's decision but is not able to directly observe the latent cognitive state, including the engagement state and adherence (see Figure \ref{fig:diagram} for a the graphical model). The objective of the AI assistant in the POMDP formulation is to maximize the probability of human correctness over time, with rewards based on the correctness of the human's decision. 

The POMDP is defined by the tuple \( \langle S, A, T, R, O, Z, \gamma \rangle \), where:

\begin{itemize}
    \item \textbf{State Space (\( S \))}: The state includes both observable and latent variables describing the system. In this framework, \( S = \langle c, y, z, \theta \rangle \), where \( c \) represents the task context, \( y \) the human decision, \( z \) adherence to AI advice, and \( \theta \) the engagement state.
    \item \textbf{Action Space (\( A \))}: The AI can either provide advice (\( a = \text{on} \)) or withhold advice (\( a = \text{off} \)).
    \item \textbf{Transition Model (\( T(s, a, s') \))}: Defines the probability of transitioning to a new state \( s' \) given the current state \( s \) and action \( a \). State transitions depend on the dynamics of engagement and decision-making, governed by Equation~\ref{eq:thetaupdate}.
    \item \textbf{Reward Function (\( R(s, a) \))}: Rewards are based solely on the correctness of the human decision:
    \begin{equation}
    R(s, a) = 
    \begin{cases} 
    r_{\text{correct}} & \text{if } y = \text{correct}, \\
    r_{\text{incorrect}} & \text{if } y = \text{incorrect}.
    \end{cases}
    \end{equation}
    Here, \( r_{\text{correct}} > r_{\text{incorrect}} \). The probability of \( y = \text{correct} \) depends on the human's adherence (\( z \)) and the AI’s action (\( a \)), as described in Equations~\ref{eq:withoutai} and~\ref{eq:withai}.
    \item \textbf{Observation Space (\( O \))}: The AI observes a subset of the state, specifically the task context \( c \) and the human decision \( y \), \( O = \langle c, y \rangle \).
    \item \textbf{Observation Model (\( Z(o \mid s', a) \))}: Specifies the probability of observing \( o \) given the resulting state \( s' \) and action \( a \). This accounts for partial observability of the human's cognitive state.
    \item \textbf{Discount Factor (\( \gamma \))}: A scalar \( 0 \leq \gamma \leq 1 \) that prioritizes long-term rewards over immediate outcomes.
\end{itemize}

The objective of the POMDP framework is to compute an optimal policy \( \pi(a \mid o) \), which determines the probability of taking action \( a \) given observation \( o \). This policy maximizes the expected cumulative discounted reward:

\begin{equation}
\mathbb{E}\left[\sum_{t=0}^\infty \gamma^t R(s_t, a_t)\right].
\end{equation}

\begin{figure*}[htb]
    \centering
    \includegraphics[width=\linewidth,
                     trim={0cm 0.0cm 0cm 0.0cm},
                     clip]{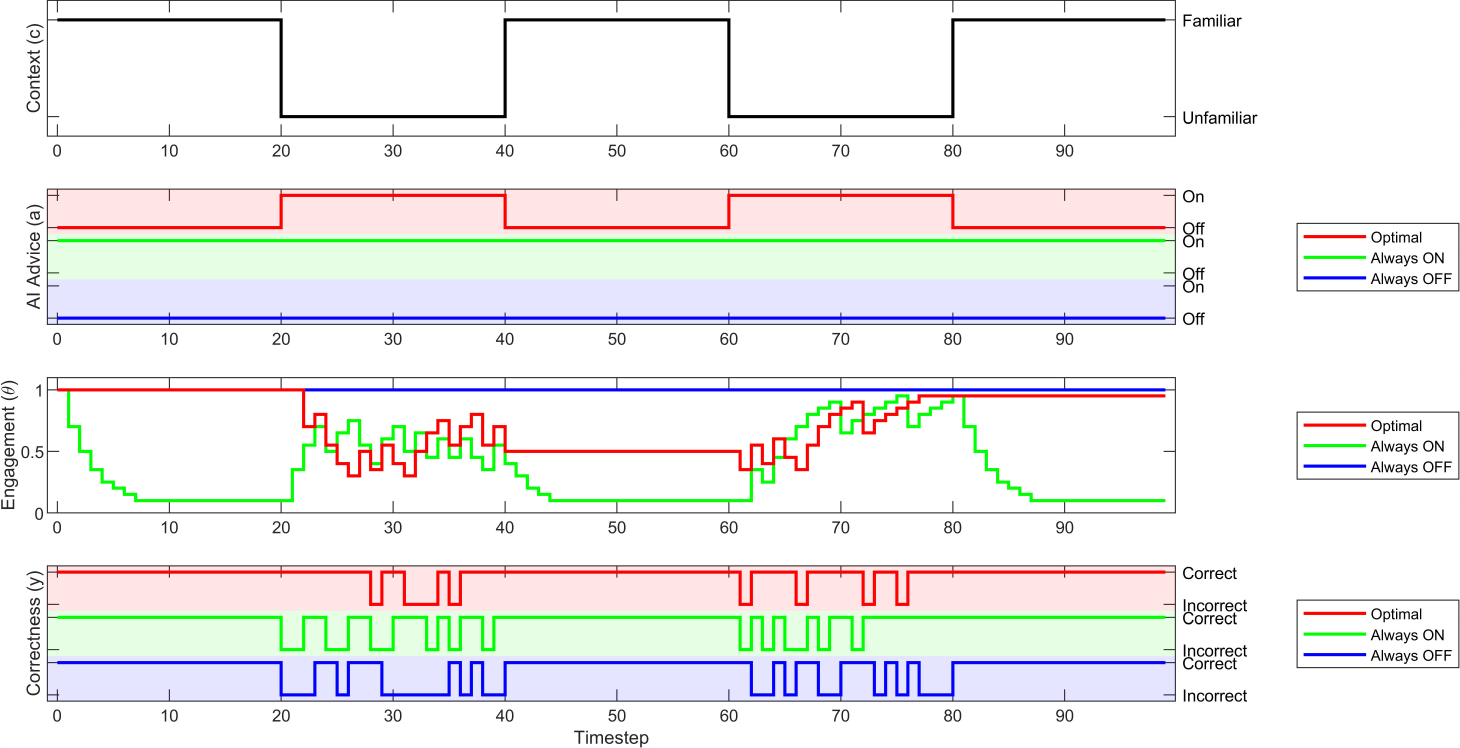}
    \caption{Illustrative results of three different AI timing strategies in an alternating task context environment. In the environment, the task context (top panel) alternates between familiar and unfamiliar contexts (represented by high and low values). The next three panels show the AI advice (on or off), the latent engagement state not observable to the AI, and the human correctness at every step.}
    \label{fig:toyresults}
\end{figure*}

\subsection{Estimating POMDP policy}
We follow a simple procedure to approximate the optimal POMPDP policy. First, we discretize the continuous \( \theta \) variable representing engagement into \( K \) bins between 0 and 1 (we used \(K=20\) in our simulations). As a result, the entire state space consists of discrete variables, which simplifies the estimation problem. 

Next, we adopted a forward-search procedure to approximate the optimal POMDP policy. Starting from a current belief state, the method enumerates each available action, simulates possible observation outcomes, and updates the belief accordingly. At each simulated step, it accumulates immediate rewards and discounts future returns, iterating this process recursively up to a defined horizon. Once the expected value is computed for each action, the procedure selects the action that maximizes this value. Repeating this across different belief states yields a policy that accounts for uncertainty in both context and correctness observations.

\section{Results}
\subsection{Model Predictions in a Simulation Environment}
We now demonstrate how the model operates in a controlled scenario by comparing three AI timing policies over 100 time steps. In this setup, the task context alternates between high and low familiarity every 20 steps (see Figure \ref{fig:toyresults}). We compare:
\begin{itemize}
    \item \textbf{Optimal (POMDP)}: A policy derived from the POMDP framework,
    \item \textbf{Always ON}: AI advice is given at every time step,
    \item \textbf{Always OFF}: AI advice is never provided.
\end{itemize}

The simulation uses Equations~(\ref{eq:withoutai})--(\ref{eq:thetaupdate}) to track human decision outcomes and engagement changes. Parameters are set to 
$\alpha_L = 0.3, \alpha_H = 1, \eta = 0.3, r_{\text{correct}} = 1, r_{\text{incorrect}} = 0, \gamma = 0.95,$ and $\phi = 0.1$. The model's initial belief state is initialized at maximum engagement: $p(\theta_0=1)=1$. the resulting approximate policy for the POMDP model does not assume prior knowledge of the regular switching interval.

Overall, the POMDP-based optimal timing policy yields 91\% correct decisions, outperforming both the always-on (85\%) and always-off (77\%) policies. In the optimal policy condition, the AI refrains from giving advice in high-familiarity contexts, preserving the user’s engagement. In contrast, the always-on policy leads to decreases in engagement by offering frequent and sometimes redundant assistance, which leads to lower adherence during challenging contexts. Although the always-off policy avoids over-advising, it never provides the human the benefit from the AI’s greater accuracy in difficult contexts, resulting in the lowest overall performance.

 \begin{table}[htb]
  \caption{Simulated human accuracy across systematic changes to model parameters while keeping other model parameters constant. Values marked with * represent default parameter values. Values between parentheses represent the probability that the optimal policy will turn on the AI advice. Boldface values represent the highest accuracy achieved for each parameter setting.}
  \label{tab:freq}
  \centering
  \begin{tabular}{lccc}
  \addlinespace 
    \toprule
  & \multicolumn{3}{c}{Policy}\\
  \cline{2-4}
  & Optimal & Always & Always \\
Variable  & Timing & On & Off \\
    \midrule
$\alpha_L$=0.00	& \textbf{92.8\%} (0.50)	& 87.5\%	& 50.5\%\\ 
$\alpha_L$=0.30*& \textbf{86.6\%} (0.50)	& 83.6\%	& 65.1\%\\ 
$\alpha_L$=0.50	& \textbf{86.2\%} (0.50)	& 84.8\%	& 74.9\%\\ 
$\alpha_L$=0.70	& \textbf{89.4\%} (0.50)	& 88.9\%	& 85.0\%\\ 
\addlinespace 
$\alpha_H$=0.70	& \textbf{76.8\%} (1.00)	& \textbf{76.8\%}	& 50.1\%\\ 
$\alpha_H$=0.80	& \textbf{78.2\%} (1.00)	& \textbf{78.2\%}	& 55.1\%\\ 
$\alpha_H$=0.90	& \textbf{82.5\%} (0.50)	& 80.6\%	& 60.1\%\\ 
$\alpha_H$=1.00*& \textbf{86.6\%} (0.50)	& 83.6\%	& 65.1\%\\ 
\addlinespace 
$\eta$=0.10	& \textbf{87.0\%} (0.50)	& 80.9\%	& 65.1\%\\ 
$\eta$=0.30*& \textbf{86.6\%} (0.50)	& 83.6\%	& 65.1\%\\ 
$\eta$=0.50	& \textbf{86.4\%} (0.50)	& 84.9\%	& 65.1\%\\ 
\addlinespace 
$\phi$=0.05	& \textbf{87.6\%} (0.51)	& 85.5\%	& 64.1\%\\ 
$\phi$=0.10*& \textbf{86.6\%} (0.50)	& 83.6\%	& 65.1\%\\ 
$\phi$=0.30	& \textbf{81.7\%} (0.50)	& 79.0\%	& 64.9\%\\ 
$\phi$=0.50	& \textbf{77.1\%} (0.49)	& 76.9\%	& 65.3\%\\ 
\addlinespace
  \bottomrule
\end{tabular}
\label{tab:paramresults}
\end{table}

\subsection{Optimal Policies Across Environments}
We now examine how variations in each parameter affect the optimal timing policy relative to simpler always-on or always-off baselines. Table~\ref{tab:paramresults} reports simulated human accuracy when we systematically adjust one parameter at a time while keeping others constant. Overall, the optimal timing strategy outperforms the baselines in most scenarios, but its advantage narrows under certain conditions.

\paragraph{Human Accuracy in Low-Familiarity Contexts (\(\alpha_L\)).}
As \(\alpha_L\) increases, the user requires less help in challenging contexts, reducing the added value of AI intervention. Consequently, the gap between the optimal timing and always-on policies shrinks.

\paragraph{Human Accuracy in High-Familiarity Contexts (\(\alpha_H\)).}
When \(\alpha_H\) decreases, users struggle more in tasks they would otherwise find easy. This slightly diminishes the benefit of withholding advice, as these contexts are no longer so reliably correct without AI support.

\paragraph{Engagement Update Rate (\(\eta\)).}
A higher \(\eta\) enables the user to recover from low engagement more quickly, causing the penalties of frequent advice to be shorter-lived. Therefore, the difference between the optimal policy and the always-on policy is less pronounced.

\paragraph{Context Switching (\(\phi\)).}
Our initial expectation was that slower context switching (i.e., smaller \(\phi\)) would give the AI more stable windows in which to adopt a well-timed advising strategy, thereby amplifying the advantages of adaptive assistance. However, the results in Table~\ref{tab:paramresults} do not provide a clear confirmation of this hypothesis. In practice, both very slow and relatively faster context switching show minimal impact on final accuracy. One possible explanation is that other factors—such as human engagement dynamics or the magnitude of \(\alpha_L\) and \(\alpha_H\)—dominate the overall performance, overshadowing the effect of \(\phi\). 

Overall, the effectiveness of strategic advising depends on user capabilities (\(\alpha_L,\alpha_H\)), the dynamics of engagement (\(\eta\)), and to a lesser extent how often the environment changes (\(\phi\)). In certain edge cases, simpler advice strategies approximate the performance of the optimal policy, but under typical conditions, adaptive advice offers a clear advantage.

\section{Discussion}
Our simulations demonstrate that an adaptive, POMDP-based approach to AI timing consistently outperforms both an ``always-on'' strategy and an ``always-off'' strategy. By selectively offering advice only when the user is likely to benefit most, the system preserves user engagement for challenging contexts and avoids overwhelming the user with unnecessary alerts. When task familiarity fluctuates over time, this adaptive policy results in higher long-term accuracy and maintains higher levels of engagement. Moreover, the exploration of the model's parameter space shows that the exact advantage of adaptive advising depends on user-specific factors—such as baseline ability and the rate at which engagement shifts—highlighting the necessity of tailoring communication strategies to different usage scenarios.

Our modeling approach has conceptual parallels with theory of mind reasoning, where the AI draws inferences about the user's internal cognitive states. By tracking engagement levels and inferring whether advice is useful or redundant, the model effectively simulates how the user perceives, values, and acts on the system's recommendations. This perspective aligns with recent work that incorporates theory of mind in AI systems \cite{rabinowitz2018machine,nguyen2022theory}, enabling adaptive behaviors that anticipate users' beliefs, intentions, or knowledge gaps. In our framework, the AI anticipates the changes in the user's engagement on the basis of the helpfulness of the AI guidance. 

The computational framework also shares conceptual parallels with the ``off-switch problem'' \cite{hadfield2017off}, in that both address scenarios where AI systems must decide when and how to limit their own intervention to achieve better long-term outcomes. In the off-switch problem, the focus is on designing AI systems that allow humans to override or deactivate them without undermining the AI's ability to pursue its objectives. Similarly, our framework examines how AI systems can strategically ``self-restrain'' by withholding assistance in familiar or redundant situations to avoid diminishing user engagement. While the off-switch problem primarily concerns the explicit relinquishing of control, our framework extends this idea to subtler forms of disengagement, such as adapting advising strategies based on the evolving attentional state of the user. Both approaches emphasize the importance of aligning AI behavior with human intentions and long-term collaboration.

The modeling framework can be extended in a number of ways. First, the model could allow for a wider variety of human engagement dynamics, ranging from permanent disengagement when attention falls below a threshold to adherence in any scenario involving uncertainty. Another promising avenue is to adapt the framework to individual differences by learning how each user responds to redundant AI interventions and by identifying which contexts are most likely to trigger disengagement. Over time, the model could build personalized engagement profiles and recognize when a user has mastered certain tasks and adjust the frequency of assistance accordingly.

\section{Conclusion}
In conclusion, by integrating communication strategies into AI system design, long-term user engagement and efficacy can be enhanced. By focusing on when to provide assistance, our framework helps AI systems avoid overwhelming users with unnecessary advice. Our findings emphasize that strategic silence can be just as valuable as offering guidance.


\bibliographystyle{apacite}

\setlength{\bibleftmargin}{.125in}
\setlength{\bibindent}{-\bibleftmargin}

\bibliography{references}

\end{document}